\documentclass[12pt]{iopart}

\usepackage{iopams}
\usepackage{graphicx, subfigure}
\usepackage{cite}
\usepackage{latexsym}
\usepackage{color}
\begin{document}

\title[Measurement of $f(v_\|, v_\perp)$ by tomographic inversion of FIDA spectra]{Measurement of a 2D fast-ion velocity distribution function by tomographic inversion of fast-ion D-alpha spectra}

\author{M~Salewski$^1$, B~Geiger$^2$, A~S~Jacobsen$^1$, M~Garc\'ia-Mu\~noz$^3$, W~W~Heidbrink$^4$, S~B~Korsholm$^1$, F~Leipold$^1$, J~Madsen$^1$, D~Moseev$^2$, S~K~Nielsen$^1$, J~Rasmussen$^1$, M~Stejner$^1$, G~Tardini$^2$, M~Weiland$^2$ and the ASDEX Upgrade team$^2$}

\address{$^1$ Association Euratom - DTU, Technical University of Denmark, Department of Physics, DTU Ris{\o}~Campus, DK-4000 Roskilde, Denmark\\
$^2$ Association Euratom - Max-Planck-Institut f\"ur Plasmaphysik, D-85748 Garching, Germany\\
$^3$ University of Seville, Department of Physics, Seville, Spain\\
$^4$ University of California, Irvine, California 92697, USA}

\ead{msal@fysik.dtu.dk}

\begin{abstract}
We present the first measurement of a local fast-ion 2D velocity distribution function $f(v_\|, v_\perp)$. To this end, we heated a plasma in ASDEX Upgrade by neutral beam injection and measured spectra of fast-ion D$_\alpha$ (FIDA) light from the plasma center in three views simultaneously. The measured spectra agree very well with synthetic spectra calculated from a TRANSP/NUBEAM simulation. Based on the measured FIDA spectra alone, we infer $f(v_\|, v_\perp)$ by tomographic inversion. Salient features of our measurement of $f(v_\|, v_\perp)$ agree reasonably well with the simulation: the measured as well as the simulated $f(v_\|, v_\perp)$ are lopsided towards negative velocities parallel to the magnetic field, and they have similar shapes. Further, the peaks in the simulation of $f(v_\|, v_\perp)$ at full and half injection energies of the neutral beam also appear in the measurement at similar velocity-space locations. We expect that we can measure spectra in up to seven views simultaneously in the next ASDEX Upgrade campaign which would further improve measurements of $f(v_\|, v_\perp)$ by tomographic inversion.
\end{abstract}

\maketitle
\section{Introduction}
\label{sec:intro}
The fast-ion phase-space distribution function is often the key to understanding many aspects of plasma behaviour but it can only be incompletely diagnosed. Here we discuss fast-ion D$_\alpha$ (FIDA) spectroscopy which measures spectra of D$_\alpha$ light at large Doppler shifts \cite{Heidbrink2010}. FIDA spectra are 1D functions of the 3D fast-ion velocity distribution function in small measurement volumes. In strongly magnetized plasmas, the 3D velocity distribution function can be regarded as 2D by decoupling the fast, quasi-periodic gyro-motion from the drift motion \cite{Alfven1940}. Hence we consider local 2D fast-ion velocity distribution functions $f(v_\|, v_\perp)$ where $v_\|$ and $v_\perp$ are velocities parallel and perpendicular to the magnetic field, respectively. We have recently shown theoretically that $f(v_\|, v_\perp)$ can be inferred from FIDA spectra by tomographic inversion \cite{Salewski2012}. Tomography is a standard analysis method in nuclear fusion research \cite{Ertl1996, Anton1996, Ingesson1998, Nagayama1996, Howard1996, Konoshima2001, Peterson2003, Furno2005, Bonheure2006, Svensson2008} as well as in many fields throughout physical and medical sciences \cite{A.C.KakandM.Slaney1988, G.T.Herman2009}. Fast-ion velocity-space tomography in nuclear fusion research has until now only been investigated theoretically \cite{Egedal2004, Salewski2011, Salewski2012, Salewski2013}. Here we apply this method to measure $f(v_\|, v_\perp)$ for the first time.

Our velocity-space tomography approach seeks to make up for shortcomings in conventional FIDA data analysis procedures. FIDA measurements are often compared with numerical simulations to judge if a measurement is consistent with a theoretical model or not. This is conventionally done by means of synthetic diagnostics using forward modelling in which the expected FIDA spectrum is modelled on the basis of a simulation of the fast-ion distribution, for example by the FIDASIM code \cite{Heidbrink2011} on the basis of a TRANSP/NUBEAM simulation \cite{Pankin2004}. If the FIDA measurements agree with synthetic FIDA measurements \cite{Heidbrink2004, Heidbrink2007, Luo2007a, Muscatello2010, vanZeeland2009, Park2009, Geiger2011, Grierson2012, Grierson2012a, Pace2013}, it is argued that $f(v_\|, v_\perp)$ in the experiment corresponds to the simulation and that the fast-ion behavior therefore is understood -- at least in the interrogation regions of the measurements. However, if they disagree \cite{Heidbrink2009a, Heidbrink2007a, Heidbrink2009, Heidbrink2008, Gorelenkov2009, Liu2010, Garcia-Munoz2011,  Heidbrink2012, Heidbrink2012a, Michael2013, Pace2011a, Muscatello2012}, it is unclear how the experimental and simulated functions $f(v_\|, v_\perp)$ are different. In this case, we would rather solve the inverse problem to know what the measured FIDA spectra imply about $f(v_\|, v_\perp)$. The tomographic inversion of the FIDA spectra is a solution to this inverse problem and provides a measurement of $f(v_\|, v_\perp)$. We could then examine in which parts of 2D velocity space the simulation and the measurement disagree. 

Nevertheless, here we chose a discharge with very good agreement between measured and synthetic spectra so that the TRANSP/NUBEAM simulation should be a good model for the discharge and should resemble our measurement of $f(v_\|, v_\perp)$. In section~\ref{sec:FIDA} we discuss the experimental conditions and the three-view FIDA diagnostic. In section~\ref{sec:algorithm} we briefly review the inversion method. We demonstrate in section~\ref{sec:results} that salient features of our measurement of $f(v_\|, v_\perp)$ agree with the simulation. Lastly, in section~\ref{sec:discussion} we discuss the potential of tomographic inversion of FIDA and other measurements, and we draw conclusions in section~\ref{sec:conclusions}.

\section{Three-view FIDA measurements}
\label{sec:FIDA}
FIDA measurements at ASDEX Upgrade were described in references \cite{Geiger2011, Geiger2013}. For our experiment in discharge 29578, we selected a plasma scenario with very low plasma density of about $n=1.8 \times 10^{19}$~m$^{-3}$ in the plasma center in order to limit bremsstrahlung and the very bright so-called passive D$_\alpha$ light from excited deuterium at the plasma edge \cite{Geiger2013}. The discharge in deuterium was heated by the co-current neutral beam injection (NBI) source Q3 that was switched on 70~ms before the measurement. NBI Q3 has an injection energy of 60~keV and a power of 2.5~MW. The toroidal magnetic field was $B_t=1.8$~T, and the current was $I_p=0.8$~MA. The plasma was in L-mode at the time of our measurements.

\begin{figure}[htbp]
\centering
    \includegraphics[width=120mm]{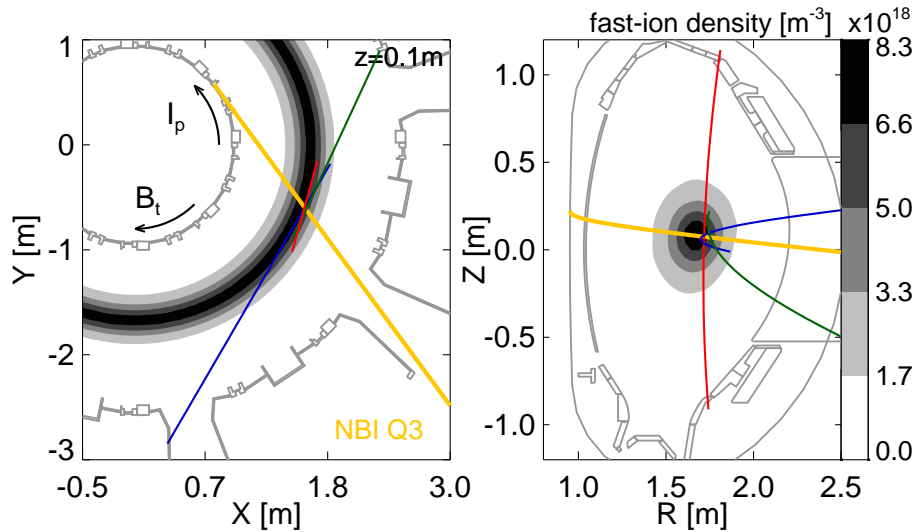}
\caption{The geometry of the three FIDA lines-of-sight (red: $\phi=69^\circ$, $R=1.720$~m; green: $\phi=156^\circ$, $R=1.749$~m; blue: $\phi=12^\circ$, $R=1.728$~m) and NBI Q3 (yellow) are shown in a toroidal view to the left and in a poloidal view to the right. The FIDA measurements are localized where NBI Q3 intersects the lines-of-sight. The fast-ion density from a TRANSP/NUBEAM simulation is illustrated in grey. The directions of the plasma current and the toroidal magnetic field are indicated in the left figure.} \label{fig:geometry}
\end{figure}

Figure~\ref{fig:geometry} illustrates the geometry of the three-view FIDA measurements in a toroidal and a poloidal view. FIDA light is generated along the path of the neutral beam where many neutrals can undergo charge exchange reactions with fast ions. FIDA measurement volumes are located at the intersections of the lines-of-sight and the path of NBI Q3. Here we choose the lines-of-sight such that the measurement volumes in each view are very similar. The spatial resolution of the three FIDA views is about 3~cm to 6~cm, and the centers of the three measurement volumes are within 3~cm of each other. The three FIDA views therefore observe approximately the same spatial volume in the plasma center.  The velocity-space interrogation regions in a FIDA view are determined by the wavelength range and the viewing angle $\phi$ between the line-of-sight and the magnetic field \cite{Heidbrink2007, Heidbrink2010, Salewski2011}. The three views have viewing angles $(12^\circ, 156^\circ, 69^\circ)$ which is equivalent to $(12^\circ, 24^\circ, 69^\circ)$ since a spectrum at $\phi=24^\circ$ is a mirror image of that at $\phi=156^\circ$. The FIDA light is collected by fibers placed in the vacuum vessel and is guided to a single 180 mm Czerny-Turner-like spectrometer with a grating with 2000 lines/mm. The spectrometer is optimized for high photon throughput (f/2.8). The spectrally dispersed radiation is measured by a low-noise electron-multiplying charge-coupled device (EM-CCD) camera with 2 ms exposure time. We used a neon lamp for the wavelength calibration and the beam emission appearing in the three spectra for the absolute intensity calibration.

\begin{figure}[htbp]
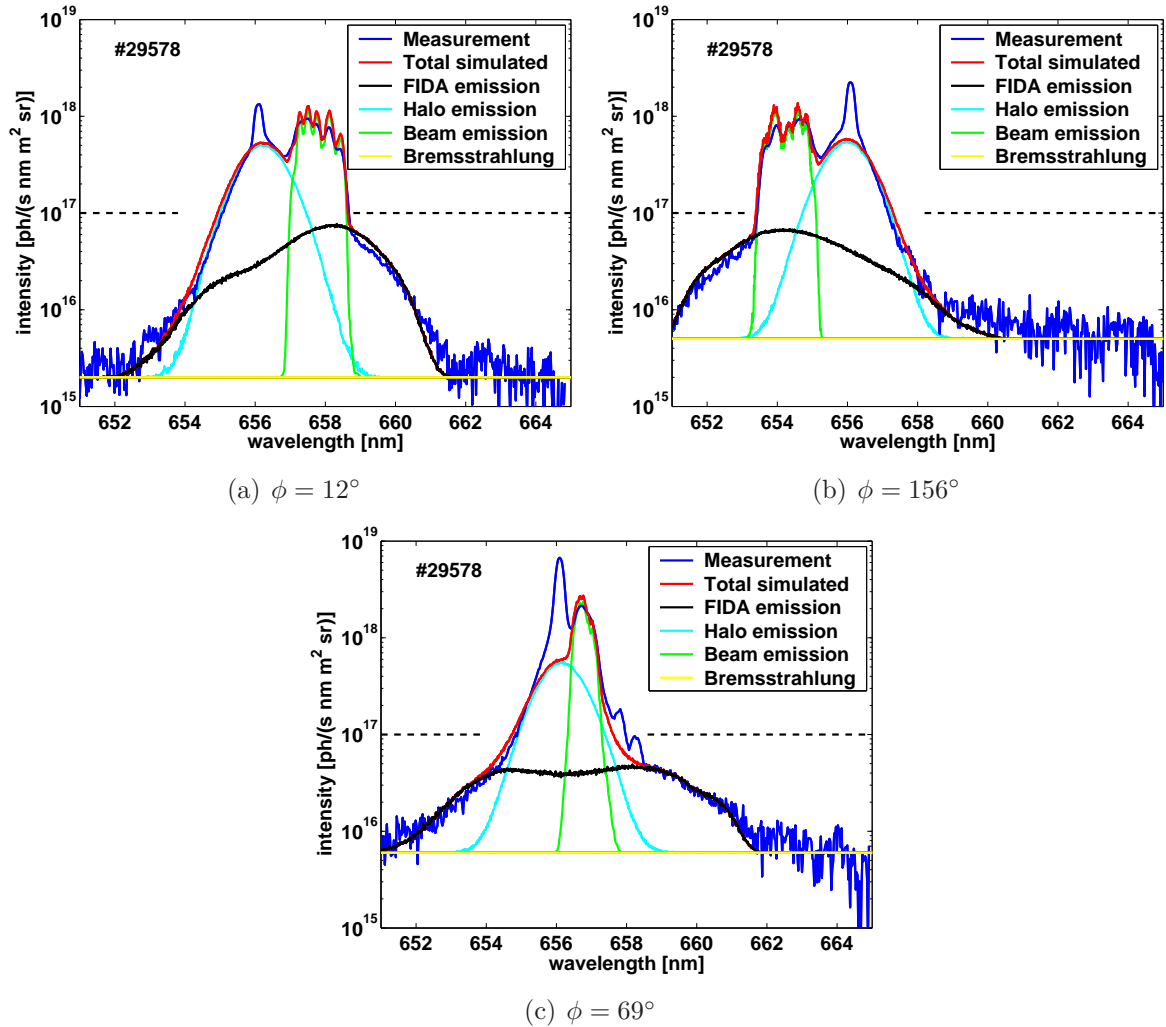

\centering
\subfigure[$\phi=12^\circ$]{%
	\includegraphics[width=76mm]{figures/29578_spectrum_ch2_NBI3.epsc}
	}
	\subfigure[$\phi=156^\circ$]{%
	\includegraphics[width=76mm]{figures/29578_spectrum_ch3_NBI3.epsc}
	}
	\subfigure[$\phi=69^\circ$]{
    \includegraphics[width=76mm]{figures/29578_spectrum_ch1_NBI3.epsc}
    }
\caption{Measured and synthetic FIDA spectra based on TRANSP/NUBEAM simulations are shown for the three FIDA views. The total simulated spectra are the sums of the contributions from FIDA, halo, and beam emissions and bremsstrahlung. The dashed lines indicate the wavelength ranges used for the inversion. The FIDA emission curves are plotted with linear axes in figure~\ref{fig:measurements_linear}.} \label{fig:measurements}
\end{figure}

\begin{figure}[htbp]
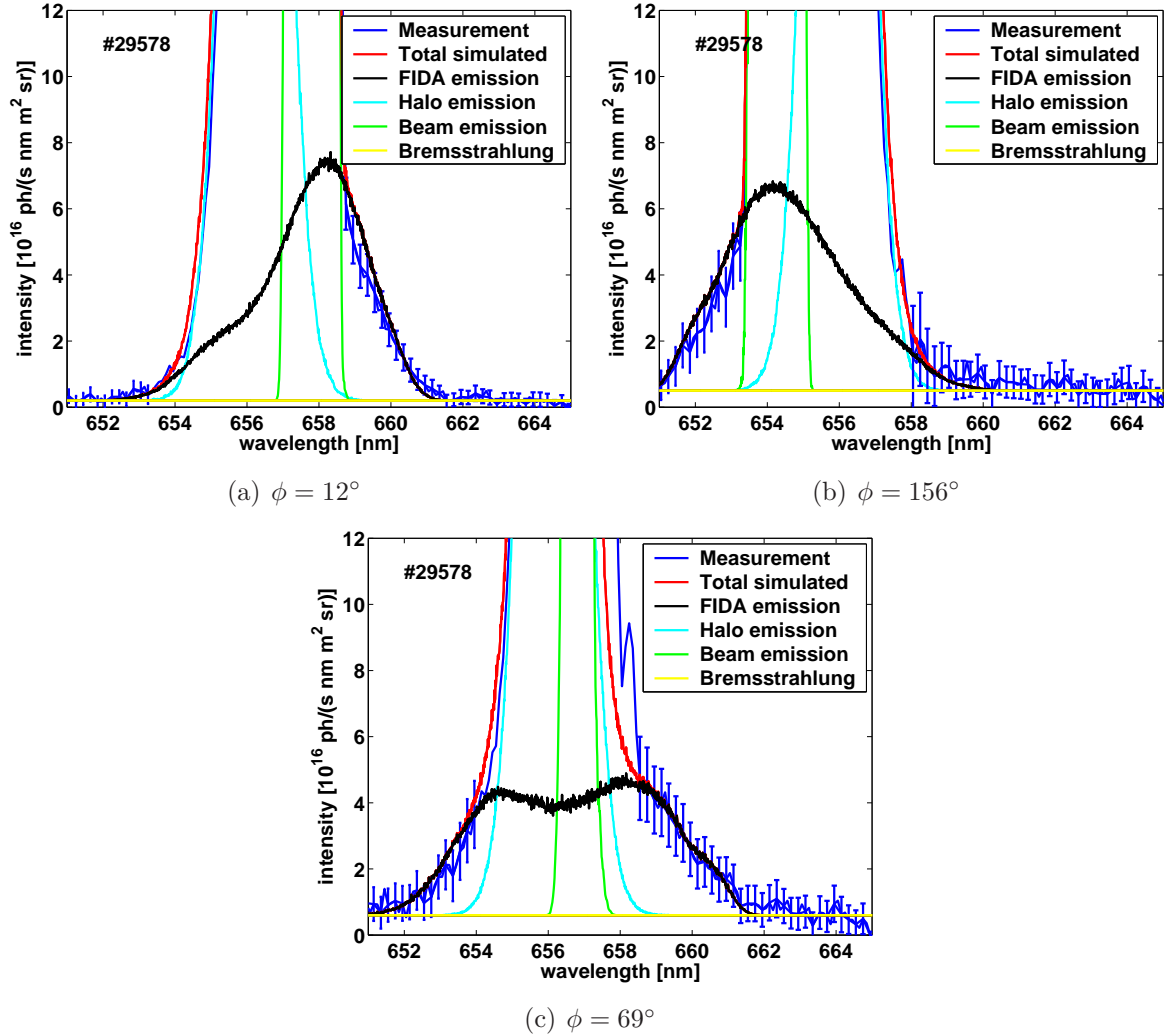

\centering
\subfigure[$\phi=12^\circ$]{%
	\includegraphics[width=76mm]{figures/29578_spectrum_ch2_NBI3_linear.epsc}
	}
	\subfigure[$\phi=156^\circ$]{%
	\includegraphics[width=76mm]{figures/29578_spectrum_ch3_NBI3_linear.epsc}
	}
	\subfigure[$\phi=69^\circ$]{
    \includegraphics[width=76mm]{figures/29578_spectrum_ch1_NBI3_linear.epsc}
    }
\caption{Measured and synthetic FIDA spectra based on TRANSP/NUBEAM simulations are shown for the three FIDA views in linear scale. The uncertainties of every second measurement are indicated. The graphs are presented with logarithmic intensities in figure~\ref{fig:measurements}.} \label{fig:measurements_linear}
\end{figure}

In figures~\ref{fig:measurements} and~\ref{fig:measurements_linear} we plot measured spectra together with synthetic spectra calculated from TRANSP/NUBEAM simulations. In figure~\ref{fig:measurements} we use logarithmic intensity axes to show the contributions due to FIDA light, halo emission, beam emission, and a flat background of bremsstrahlung. In figure~\ref{fig:measurements_linear} we focus on the FIDA contributions with uncertainties using linear intensity axes. The measured spectra are averages over three frames from t=0.861~s - 0.867~s. Hence the time resolution is 6~ms. Figure~\ref{fig:measurements} shows that the sum of these four contributions in the synthetic spectra agrees very well with the measured spectra. The levels of bremsstrahlung and passive D$_\alpha$ light were so low that we could detect red-shifted as well as blue-shifted FIDA light in all three spectra. However, line radiation from the plasma edge is not taken into account in the model and hence causes discrepancies between measurements and simulations in particular wavelength ranges. The dominating line is the cold D$_\alpha$ line at 656.1~nm in all three views. In figures~\ref{fig:measurements}(b) and (c) two carbon lines appear at 657.8~nm and 658.3~nm \cite{NIST}. Figure~\ref{fig:measurements_linear} confirms that in particular the measured and synthetic FIDA emission contributions to the spectra, which we use for the tomographic inversion, agree very well despite the detected weak MHD activity at the frequency of Alfv{\'e}n modes. The Alfv{\'e}n modes may be the reason for the slight tendency of the measured spectra to lie below the synthetic spectra, but this minor discrepancy is within the experimental uncertainty.

\section{Tomographic inversion method}
\label{sec:algorithm}
We use the inversion prescription discussed in references \cite{Salewski2012, Salewski2013}. The tomographic inversion $F^+$, i.e. our estimate of the fast-ion distribution function leading to the FIDA measurements, is given by
\begin{equation}
F^+=\hat{W}^+ \hat{G}.
\end{equation}
$\hat{G}$ is a matrix containing the FIDA measurements normalized by their uncertainties, $\hat{W}^+$ is the Moore-Penrose pseudoinverse of the transfer matrix $\hat{W}$ which is composed of likewise normalized weight functions, and $F^+$ is a matrix containing the discretized function $f(v_\|, v_\perp)$ \cite{Salewski2013}. The weight functions relate $f(v_\|, v_\perp)$ to the fast-ion measurements \cite{Heidbrink2007, Heidbrink2010, Salewski2011}. The weight functions are determined by Doppler shift and Stark splitting of the emitted D$_\alpha$ light as well as charge exchange and photon emission probabilities of fast ions based on averaged neutral densities in the measurement volume as calculated by FIDASIM. The uncertainties are here given by the diagnostic read-out-noise and by the photon noise.   $\hat{W}^+$ is found by singular value decomposition of $\hat{W}$ using the largest 65 singular values whereas the remaining singular values have been truncated. The inversion $F^+$ minimizes the least-square figure of merit $\chi^2$ in which the misfit of each measurement is normalized by its uncertainty:
\begin{equation}
\chi^2=\mid \hat{G} - \hat{W}F \mid^2.
\label{eq:chi2matrix}
\end{equation}
Based on the achievable spectral resolution of the FIDA measurements, we divide the spectral range in a FIDA view into 160 wavelength intervals. Each wavelength interval monitors a particular velocity-space region described by a weight function. As we use three FIDA views, we have in total 480 weight functions. Of these, 217 weight functions cover the velocity-space of interest and spectral ranges which are dominated by FIDA light and bremsstrahlung or just by bremsstrahlung (see figure~\ref{fig:measurements}). The tomographic information lies in the amplitudes of the FIDA light. Beam emission, halo emission, or impurity lines dominate the other spectral ranges. These contributions cannot be subtracted accurately enough from the total measured signal to extract a useful estimate for the FIDA signal, and hence we do not use these spectral ranges. We are free to choose the numerical grid to describe $f(v_\|, v_\perp)$. On the one hand, we would like to describe $f(v_\|, v_\perp)$ in high resolution and so would prefer a fine grid with many grid points. On the other hand, there is a limited number of measurements to infer useful values for $F^+$ at these grid points. If there are more grid points than measurements or even if the numbers are comparable, signatures of weight functions appear in the tomographic inversion \cite{Salewski2012}. These are systematic artefacts in the inversion which can be identified but which we avoid here. We therefore choose a discretization with $17 \times 8$ grid points in ($v_\|,v_\perp$) such that the number of grid points (136) is lower than the number of measurements (217).

\section{Tomographic inversion of three-view FIDA measurements}
\label{sec:results}
The three measured FIDA spectra in discharge 29578 agreed very well with synthetic FIDA spectra based on a TRANSP/NUBEAM simulation. Therefore the simulation should be a realistic model of $f(v_\|, v_\perp)$ in the experiment, and the measurement of $f(v_\|, v_\perp)$ by tomographic inversion should resemble the TRANSP/NUBEAM simulation. 

\begin{figure}[htbp]
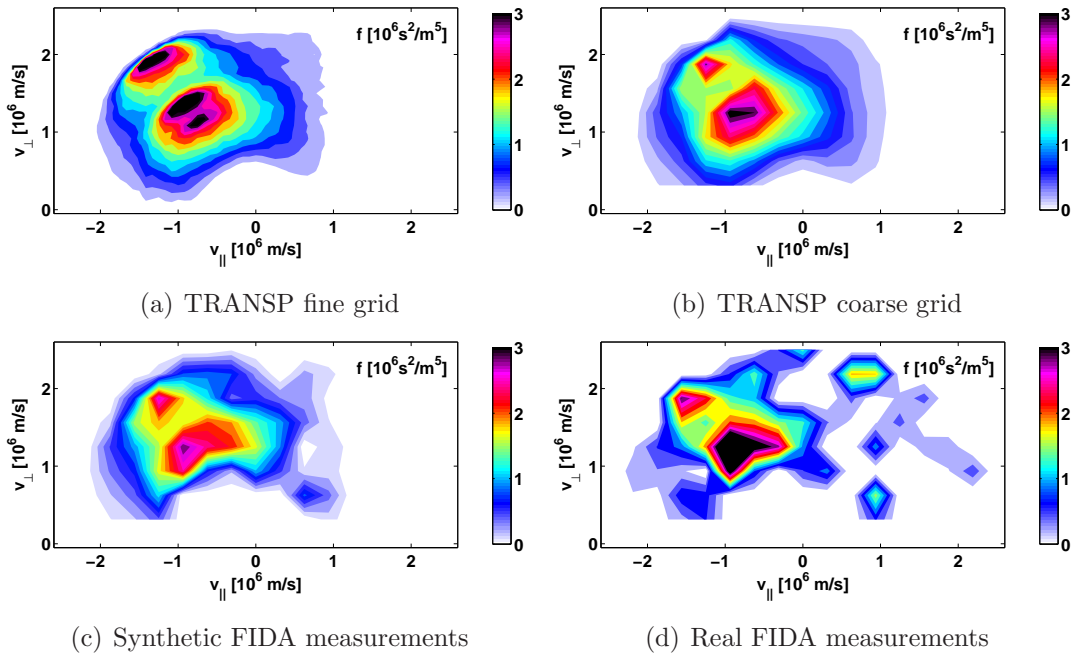

\centering
\subfigure[TRANSP fine grid]{%
	\includegraphics[width=70mm]{figures/S3fine.epsc}
	}
	\subfigure[TRANSP coarse grid]{%
	\includegraphics[width=70mm]{figures/S3.epsc}
	}
\subfigure[Synthetic FIDA measurements]{%
	\includegraphics[width=70mm]{figures/reconstruction.epsc}
	}
	\subfigure[Real FIDA measurements]{%
	\includegraphics[width=70mm]{figures/tomo.epsc}
	}
\caption{Simulation of $f(v_\|, v_\perp)$ as well as measurement of $f(v_\|, v_\perp)$ by tomographic inversion. (a) TRANSP/NUBEAM simulation discretised by $61 \times 30$ grid points. (b) TRANSP/NUBEAM simulation discretised by $17 \times 8$ grid points. (c) Inversion of synthetic FIDA spectra based on the TRANSP/NUBEAM simulation. (d) Measurement of $f(v_\|, v_\perp)$ by tomographic inversion of FIDA spectra.} \label{fig:fvpavpe}
\end{figure}

Figure~\ref{fig:fvpavpe}a shows the TRANSP/NUBEAM simulation of $f(v_\|, v_\perp)$ discretized on $61 \times 30$ grid points in ($v_\|,v_\perp$). This resolution is high enough to capture essential features of $f(v_\|, v_\perp)$. The function shows three peaks at full, half and one-third beam injection energies of NBI Q3 ($E=60$~keV, $E/2=30$~keV and $E/3=20$~keV). As NBI Q3 is injected against the direction of the magnetic field (co-current), these beam injection peaks have negative $v_\|$ coordinates. Hence the function is lopsided towards negative $v_\|$. We choose to calculate the inversion of the FIDA measurements on a coarser grid with $17 \times 8$ grid points in ($v_\|,v_\perp$) so that we infer 136 values from 217 measurements. In figure~\ref{fig:fvpavpe}b we interpolate the TRANSP/NUBEAM simulation to the coarse grid. On this grid the peaks at $E/2$ and $E/3$ merge to become a single dominant peak in $f(v_\|, v_\perp)$ due to the lower resolution. This interpolation in figure~\ref{fig:fvpavpe}b shows how well a measurement of $f(v_\|, v_\perp)$ could at best resemble the simulation in figure~\ref{fig:fvpavpe}a for our choice of grid for the inversion, given that the TRANSP simulation is a good model for $f(v_\|, v_\perp)$ in discharge 29578 and that the measurements are noise-free and cover the entire 2D velocity-space.

Figure~\ref{fig:fvpavpe}c shows a tomographic inversion of the synthetic spectra based on the TRANSP/NUBEAM simulation. We use only synthetic spectral data from experimentally accessible wavelength intervals for our three-view FIDA measurements. This inversion is an idealized prediction for the measurement of $f(v_\|, v_\perp)$ in figure~\ref{fig:fvpavpe}d and represents how well our measurement of $f(v_\|, v_\perp)$ with our particular three-view FIDA instrument could at best resemble the simulation in figure~\ref{fig:fvpavpe}a, given noise-free measurements and that the TRANSP simulation is a good model for $f(v_\|, v_\perp)$ in discharge 29578. As expected, the inversion from synthetic spectra resembles the TRANSP simulation very well \cite{Salewski2012}. 

In figure~\ref{fig:fvpavpe}d we present the measurement of $f(v_\|, v_\perp)$ by tomographic inversion of FIDA spectra. The TRANSP/NUBEAM simulation and the measurement of $f(v_\|, v_\perp)$ agree reasonably well. The shapes of the measured and simulated functions $f(v_\|, v_\perp)$ are similar, and the measured function $f(v_\|, v_\perp)$ is also lopsided towards negative $v_\|$. The beam injection peaks at $E$ and $E/2$ appear in the measurement very close to the velocity-space positions in the simulation. The differences between figures~\ref{fig:fvpavpe}c and \ref{fig:fvpavpe}d originate from the slight systematic differences between the measured and the synthetic FIDA spectra and from noise in the measurements. We interpret the jitter appearing in figure~\ref{fig:fvpavpe}d not to be physical but rather to be artefacts generated by noise.


To conclude, we find remarkable agreement between the measurement and the simulation of $f(v_\|, v_\perp)$. The overall shape of $f(v_\|, v_\perp)$ including the positions of the beam injection peaks can be revealed by tomographic inversion but not by conventional inspection of the FIDA spectra in figure~\ref{fig:measurements}.

\section{Discussion}
\label{sec:discussion}
In this section we suggest how to improve tomographic inversions further and discuss the potential of tomographic inversions to study physical phenomena in plasmas. In the experiment reported here, three FIDA spectra with different viewing angles were simultaneously measured and inverted. Two additional FIDA views should become available in the next ASDEX Upgrade campaign. Moreover, we can increase the signal-to-noise ratio of the FIDA measurements. Here we allowed the cold D$_\alpha$ line to enter the spectrometer together with the FIDA signal. It will benefit the signal-to-noise ratio to block the cold D$_\alpha$ line as then the EM-CCD camera can be operated at higher gain without saturation.

Further, our analysis method is not restricted to FIDA measurements. Likewise, 1D collective Thomson scattering (CTS) measurements can be inverted \cite{Salewski2012}, as can combinations of CTS and FIDA measurements \cite{Salewski2013}. Traditional CTS data analysis procedures also rely on synthetic diagnostics \cite {Moseev2011a, Salewski2010a}, and our velocity-space tomography approach will have the same benefits for the interpretation of CTS spectra. The next step will be to include the CTS diagnostic installed at ASDEX Upgrade \cite{Salewski2010a, Meo2008, Meo2010, Furtula2012, Furtula2012a} which we recently upgraded with a second radiometer. Hence a total of seven views should become available for a combined FIDA and CTS system in the next campaign. The additional four views would improve the inversions, in particular if they have large viewing angles to complement the viewing angles at 12$^\circ$ and 24$^\circ$. CTS measurements at very high frequency resolution, such as those demonstrated at TEXTOR \cite{Stejner2010b, Korsholm2011, Stejner2013}, are also possible at ASDEX Upgrade. High resolution CTS measurements have so far been restricted to bulk-ion measurements since the bandwidth and the bit resolution were not large enough to measure fast ions. High resolution spectra should contain a wealth of information suitable for tomographic inversion of future CTS measurements \cite{Salewski2012}. Measurements from other fast-ion diagnostics at ASDEX Upgrade could also be included in the inversion if weight functions describing these can be formulated, for example neutron spectroscopy \cite{Giacomelli2011, Tardini2012} and neutron count measurements \cite{Tardini2013}, $\gamma$~spectroscopy \cite{Nocente2012}, neutral particle analyzers \cite{Akaslompolo2010}, or fast-ion loss detectors \cite{Garcia-Munoz2007, Garcia-Munoz2011}. Lastly, we could likely achieve further improvements of the tomography method by using alternative inversion algorithms such as those in other branches of tomography \cite{Ertl1996, Anton1996, Ingesson1998, Nagayama1996, Howard1996, Konoshima2001, Peterson2003, Furno2005, Bonheure2006, Svensson2008, A.C.KakandM.Slaney1988, G.T.Herman2009}.

Several other machines have multi-view FIDA systems installed or are planning to install or upgrade multi-view FIDA systems. Similar measurements of $f(v_\|, v_\perp)$ could be done on DIII-D (three FIDA views \cite{Pace2011, Heidbrink2012}), NSTX (two FIDA views \cite{Bortolon2010}), MAST (two FIDA views \cite{Michael2013, Jones2013}), LHD (one CTS view \cite{Kubo2010, Nishiura2013} and two FIDA views \cite{Ito2010, TakafumiITOMasakiOSAKABEKatsumiIDAMikiroYOSHINUMAMasahikoKOBAYASHISadayoshiMURAKAMIMotoshiGOTOYasuhikoTAKEIRIDetlevREITER2012}) and ITER (one CTS view \cite{Salewski2009, Salewski2009a, Salewski2008, Leipold2009} and possibly charge exchange measurements of fast $\alpha$ particles \cite{Kappatou2012}).

The velocity-space tomography approach can potentially reveal new physics in cases with anomalous transport of fast ions, in particular if this transport depends on the position of the fast ions in velocity space. Our measurement of $f(v_\|, v_\perp)$ was made in a plasma with relatively weak Alfv{\'e}n modes. A TRANSP/NUBEAM simulation assuming neoclassical transport matched the FIDA measurements well, and we could demonstrate that it is possible to measure $f(v_\|, v_\perp)$.  Stronger Alfv{\'e}n eigenmodes than the ones reported here are known to affect ions in specific parts of velocity space \cite{Heidbrink1994, Zweben2000, Nabais2010, Garcia-Munoz2011, vanZeeland2011, Pace2011}. The amplitude of the anomalous transport due to resonant interaction of the modes with the fast ions depends on the position in velocity space as only ions fulfilling a resonance condition with the modes are affected. Several other types of modes also affect ions selectively in velocity space, for example sawteeth, neoclassical tearing modes, and or fishbones. Sawtooth crashes redistribute passing ions more than trapped ions \cite{Pace2011a, Muscatello2012, Nielsen2010, Nielsen2011}. Strong and coherent fast-ion losses observed in the presence of neoclassical tearing modes have shown that these modes selectively redistribute fast particles under resonance conditions \cite{Garcia-Munoz2007, Garcia-Munoz2009}. Fishbones \cite{Heidbrink1994, PerezVonThun2012} and off-axis fishbones \cite{Heidbrink2011b} have also been demonstrated to act selectively in velocity space. However, the underlying mechanisms leading to enhanced fast-ion transport in the presence of these MHD instabilities are still not understood well. Likewise, any anomalous fast-ion transport in the presence of microinstabilities is thought to be selective in velocity space and is not well understood \cite{Heidbrink2009, Pace2013, Hauff2009}. Our tomography approach can pinpoint the origin of any observed discrepancies between FIDA measurements and simulation in velocity space which has not been possible before. This can give clues to reveal the underlying mechanisms and can enable us to improve the forward modelling, either in cases where existing theory is wrong or even if there is no theory yet.

\section{Conclusions}
\label{sec:conclusions} 
We demonstrated that it is possible to measure salient features of a 2D fast-ion velocity distribution function $f(v_\|, v_\perp)$. For this purpose we measured spectra of FIDA light from the plasma center at ASDEX Upgrade simultaneously in three views and calculated a tomographic inversion of these measurements. The measured spectra agree very well with synthetic spectra based on a TRANSP/NUBEAM simulation, and the tomographic inversion therefore agrees reasonably well with the simulation. The inversion as well as the simulation are lopsided towards negative $v_\|$, and the overall shapes also agree reasonably well. The velocity-space locations of the beam injection peaks at $E$ and $E/2$ in the inversion are very close to those in the simulation. We hope to measure $f(v_\|,v_\perp)$ in seven views simultaneously in the next ASDEX Upgrade campaign, and this would further improve the diagnostic potential of tomographic inversion to measure 2D fast-ion velocity-space distribution functions $f(v_\|,v_\perp)$.

\section*{Acknowledgments}
This work, supported by the European Communities under the contract of Association between Euratom and DTU was partly carried out within the framework of the European Fusion Development Agreement. The views and opinions expressed herein do not necessarily reflect those of the European Commission. 
 
\section*{References}
\bibliography{salewski2012bibshort}



\end{document}